\documentclass[aps,prl,twocolumn,superscriptaddress,letterpaper]{revtex4-1}

\usepackage{graphicx}
\usepackage{bm}
\usepackage{amssymb}
\usepackage{color}
\usepackage{amsmath}
\usepackage{ulem}

\begin{document}

\title{Time-periodic corner states from Floquet higher-order topology}

\author{Weiwei Zhu}
\thanks{These authors contribute equally.}
\affiliation{Department of Physics, National University of Singapore, Singapore 117542, Singapore}

\author{Haoran Xue}
\thanks{These authors contribute equally.}
\affiliation{Division of Physics and Applied Physics, School of Physical and Mathematical Sciences, Nanyang Technological University,
Singapore 637371, Singapore}

\author{Jiangbin Gong}
\email{phygj@nus.edu.sg}
\affiliation{Department of Physics, National University of Singapore, Singapore 117542, Singapore}

\author{Yidong Chong}
\email{yidong@ntu.edu.sg}
\affiliation{Division of Physics and Applied Physics, School of Physical and Mathematical Sciences, Nanyang Technological University,
Singapore 637371, Singapore}
\affiliation{Centre for Disruptive Photonic Technologies, Nanyang Technological University,
Singapore 637371, Singapore}

\author{Baile Zhang}
\email{blzhang@ntu.edu.sg}
\affiliation{Division of Physics and Applied Physics, School of Physical and Mathematical Sciences, Nanyang Technological University,
Singapore 637371, Singapore}
\affiliation{Centre for Disruptive Photonic Technologies, Nanyang Technological University,
Singapore 637371, Singapore}

\maketitle

\textbf{The recent discoveries of higher-order topological insulators (HOTIs) have shifted the paradigm of topological materials, which was previously limited to topological states at boundaries of materials, to those at boundaries of boundaries, such as corners \cite{benalcazar2017a,benalcazar2017b,langbehn2017,song2017,schindler2018,ezawa2018}. So far, all HOTI realisations have assumed static equilibrium described by time-invariant Hamiltonians \cite{serra2018,peterson2018,imhof2018,noh2018,schindler2018b,xue2019,ni2019,zhang2019}, without considering time-variant or nonequilibrium properties. On the other hand, there is growing interest in nonequilibrium systems \cite{rudner2020} in which time-periodic driving, known as Floquet engineering, can induce unconventional phenomena including Floquet topological phases \cite{kitagawa2010,rudner2013,rechtsman2013,McIver2020,wintersperger2020} and time crystals \cite{zhang2017,choi2017}. Recent theories have attemped to combine Floquet engineering and HOTIs \cite{bomantara2019,rodriguez2019,seshadri2019,nag2019,peng2019b,ghosh2020,hu2020,huang2020,peng2020,chaudhary2019,zhang2020,zhu2020}, but there has thus far been no experimental realisation. Here we report on the experimental demonstration of a two-dimensional (2D) Floquet HOTI in a three-dimensional (3D) acoustic lattice, with modulation along $z$ axis serving as an effective time-dependent drive. Direct acoustic measurements reveal Floquet corner states that have time-periodic evolution, whose period can be even longer than the underlying drive, a feature previously predicted for time crystals. The Floquet corner states can exist alongside chiral edge states under topological protection, unlike previous static HOTIs. These results demonstrate the unique space-time dynamic features of Floquet higher-order topology. }

\begin{figure*}
  \centering
  \includegraphics[width=0.9\textwidth]{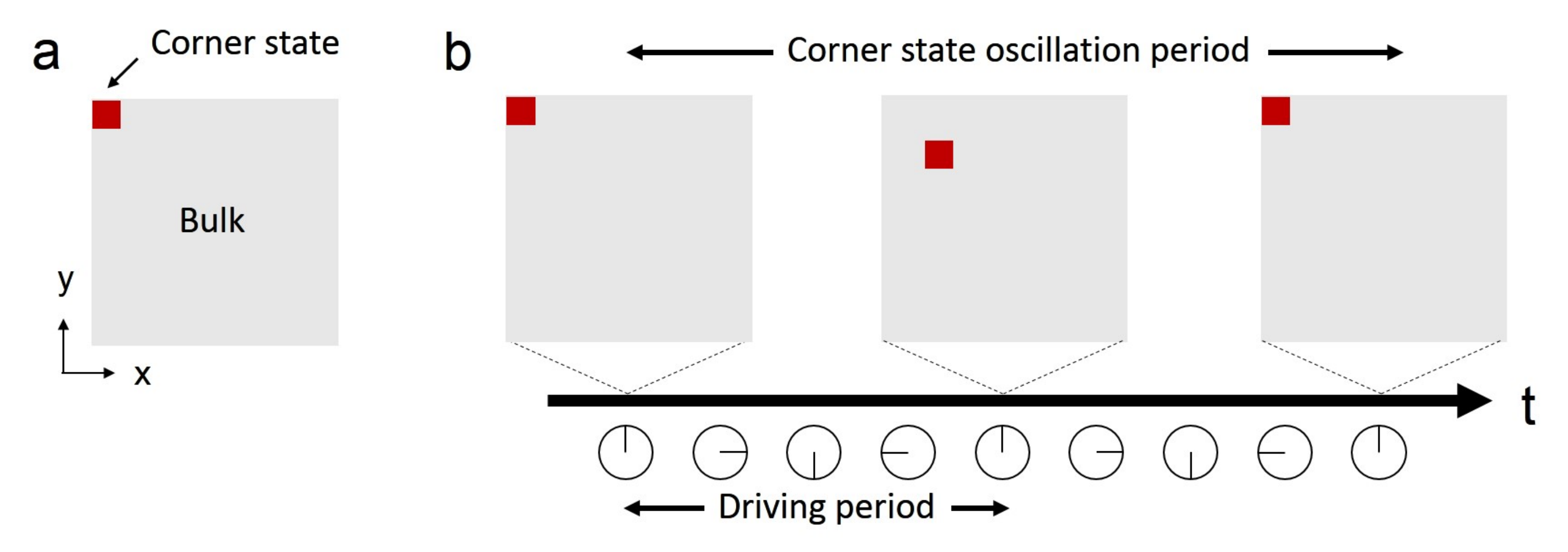}
  \caption{\textbf{Schematic comparison between static and Floquet higher-order topological insulators.} \textbf{a}, In a conventional second-order topological insulator, a corner state is localised at the corner with a time-invariant spatial distribution. \textbf{b}, In a time-periodically driven, or Floquet, second-order topological insulator, the corner state oscillates in time near the corner. The corner state oscillation period can be different from the driving period. The schematic illustrates the scenario of a doubled period. }
  \label{fig1}
\end{figure*}

\begin{figure*}
  \centering
  \includegraphics[width=0.8\textwidth]{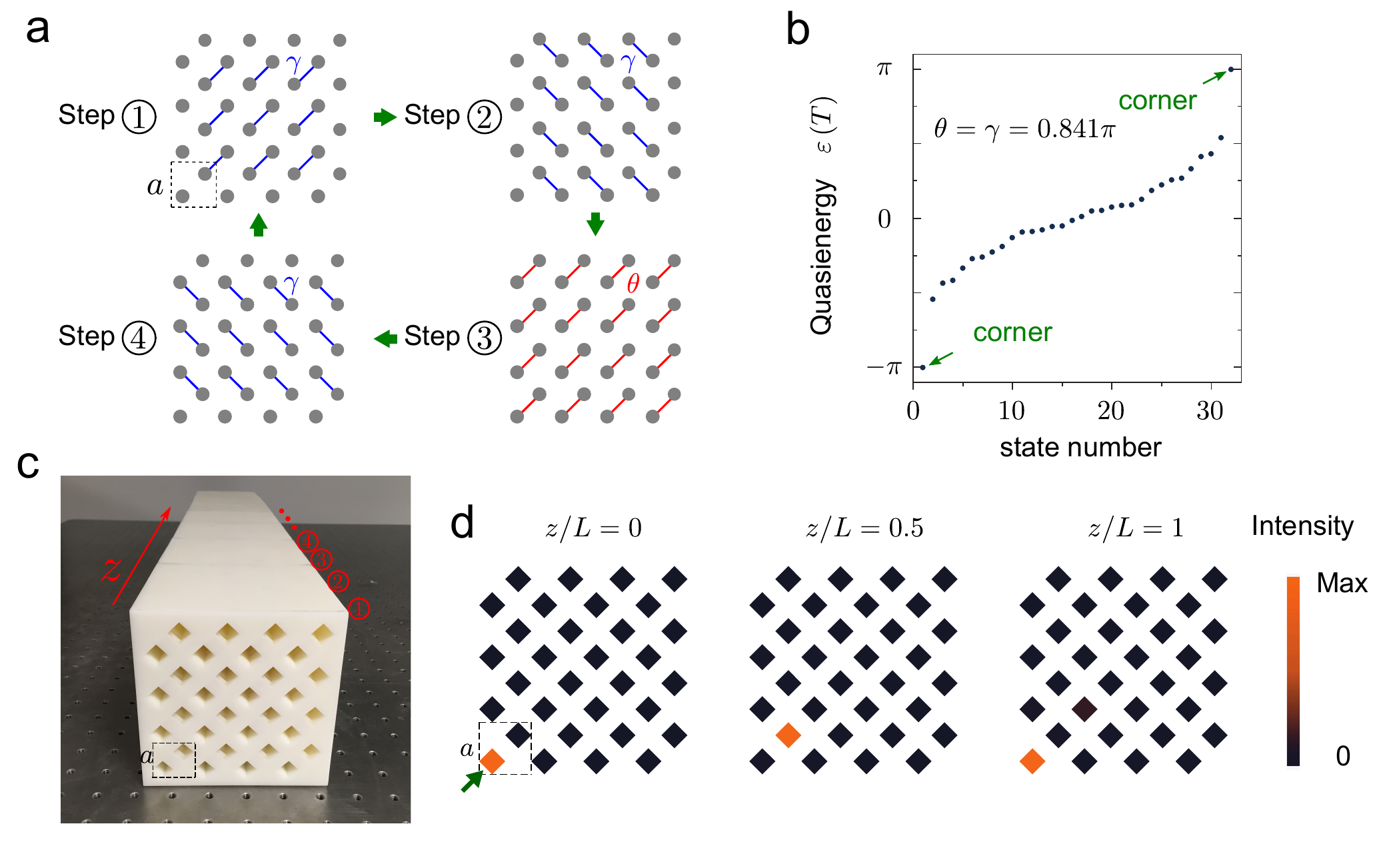}
  \caption{ \textbf{Design and construction of a Floquet higher-order topological insulator in an acoustic lattice.} \textbf{a}, Tight-binding model and driving protocol. The drive consists of four steps with equal duration $T/4$, where $T$ is the driving period. The order of the four steps is indicated by the green arrows. In each step, one lattice site couples to one of its four neighbouring sites.  The coupling strength is $\gamma$ in steps 1,2 and 4, and $\theta$ in step 3. The dotted square in step 1 indicates the unit cell. \textbf{b}, Simulated quasienergy spetrum with $\theta = \gamma = 0.841 \pi$. \textbf{c}, Photo of the fabricated acoustic structure that realises the tight-binding model in \textbf{a}. Here the $z$ axis plays the role of time. The lattice constant $a=20\sqrt{2}\,\textrm{mm}$. The coupling strength $\theta = \gamma = 0.841 \pi$ is accomplished by thin connecting waveguides that are not visible in the photo. \textbf{d}, Measured acoustic intensity distributions at different evolution distances at 8000 Hz. The green arrow indicates the excitation position at $z/L = 0$.}
  \label{fig2}
\end{figure*}

HOTIs are a class of recently discovered topological phases of matter that go beyond the framework of conventional topological physics \cite{benalcazar2017a,benalcazar2017b,langbehn2017,song2017,schindler2018,ezawa2018,serra2018,peterson2018,imhof2018,noh2018,schindler2018b,xue2019,ni2019,zhang2019}. As a typical example, a 2D second-order topological insulator---unlike a conventional 2D topological insulator that supports one-dimensional (1D) topological edge states along its edges---will instead host zero-dimensional (0D) corner states at their corners, as determined by the nontrivial higher-order bulk topology (see Fig.~1\textbf{a}). This generalised  bulk-boundary correspondence predicts the existence of topological states at lower-dimensional boundaries (e.g., corners) than conventional topological states, allowing for the topological characterisation of many materials that would have previously been considered trivial, such as twisted bilayer graphene \cite{park2019}. The identification of HOTIs has attracted great interest among fields ranging from condensed matter to photonics and acoustics. In particular, HOTIs have been realised in various classical ``metamaterial'' systems \cite{serra2018,peterson2018,imhof2018,noh2018,xue2019,ni2019,zhang2019}, due to the ease with which metamaterial properties can be tuned.

These previously-realised HOTIs have nontrivial structure only in spatial dimensions.  Yet time is another dimension that can be used to generate interesting behaviours. Time-periodic driving, or Floquet engineering, produces nonequilibrium systems whose Hamiltonians are time-periodic, i.e., $H(t+T) = H(t)$, where $T$ is the driving period. Such systems can exhibit unconventional topological phases known as Floquet topological insulators \cite{rudner2020}, which have properties that do not nonexistent in their static counterparts. For example, chiral edge states can exist in a Floquet topological insulator even though all bulk bands possess zero Chern number \cite{rudner2013,hu2015,gao2016,maczewsky2017,mukherjee2017}, in sharp constrast to standard chiral edge states. Interestingly, it has been proposed that Floquet topological states can be utilised to construct time crystals with broken discrete (or Floquet) time-translation symmetry, exhibiting period-doubling oscillation \cite{bomantara2018,bomantara2020c}, but this phenomenon has not yet been observed.  A few proposals have attempted to combine Floquet engineering with higher-order topology \cite{bomantara2019,rodriguez2019,seshadri2019,nag2019,peng2019b,ghosh2020,hu2020,huang2020,peng2020,chaudhary2019,zhang2020,zhu2020}, but none has been realised yet.

Here, we experimentally demonstrate a Floquet HOTI exhibiting topological corner states with space-time dynamics (see the schematic in Fig.~1\textbf{b}). Distinct to static HOTIs, the corner states in the Floquet HOTI can oscillate in time, with oscillation period equal to, or double of, the driving period.  The latter case is the main feature in previous proposals for discrete time crystals. We further demonstrate the coexistence of corner states and chiral edge states in a Floquet HOTI, both of which are topologically protected. This phenomenon is absent in static systems. Such corner states have no static counterpart, in the sense that the quasienergy bands have vanishing polarization and quadrupole moments, which are the usual hallmarks of trivial states in static higher-order topology. These unusual dynamic properties observed in experiment characterise the Floquet higher-order topology constructed in our study.

\begin{figure*}
  \centering
  \includegraphics[width=0.8\textwidth]{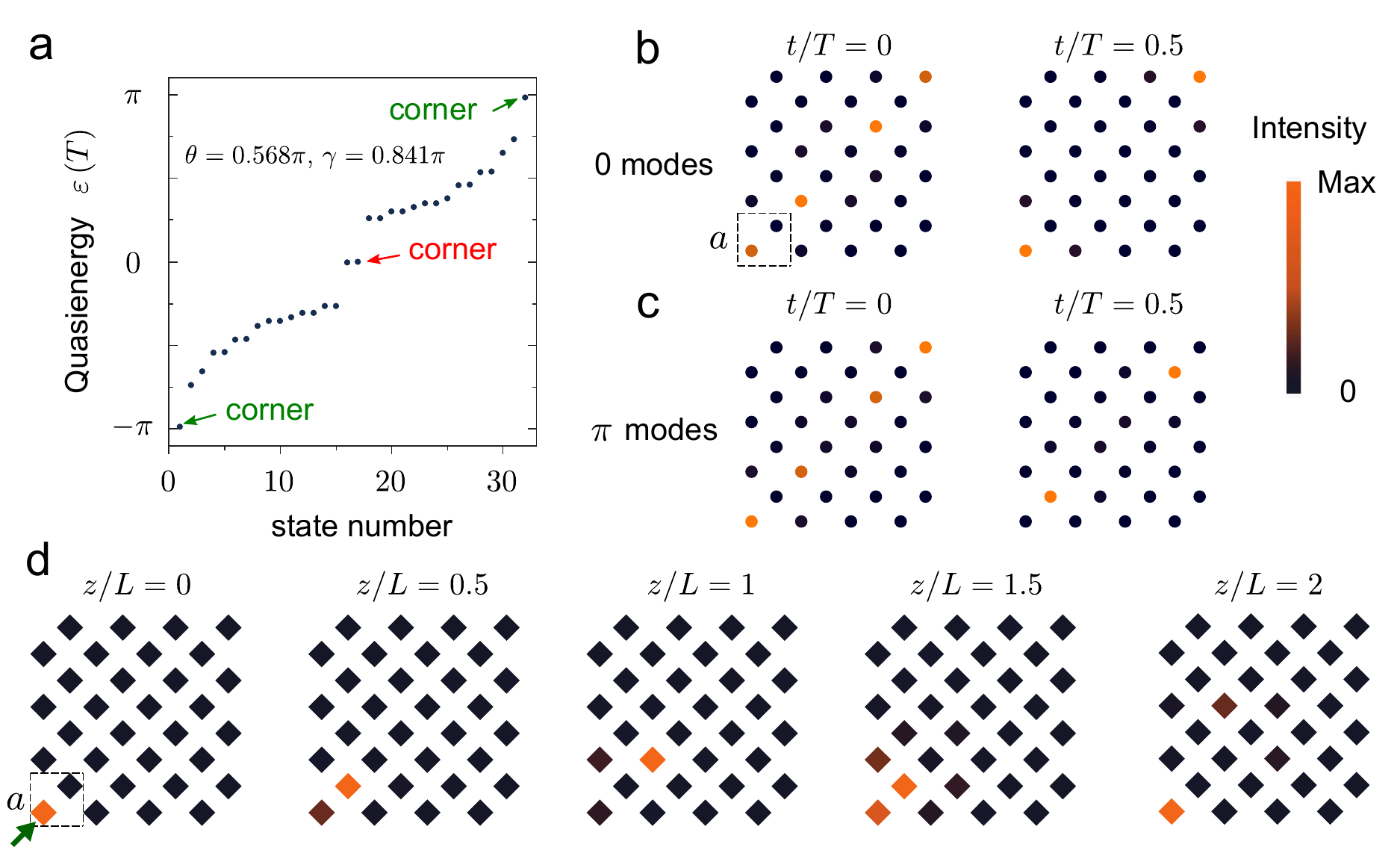}
  \caption{\textbf{Demonstration of period doubling of Floquet corner states.} \textbf{a}, Simulated quasienergy spectrum with $\theta = 0.568\pi$ and $\gamma = 0.841 \pi$. \textbf{b}, Simulated eigenmode profile for the zero corner modes. \textbf{c},  Simulated eigenmode profile for the $\pi$ corner modes.  \textbf{d}, Measured acoustic intensity distributions at different evolution distances at 8000 Hz.  The green arrow indicates the position of excitation at $z/L = 0$.}
  \label{fig3}
\end{figure*}

We start with a tight-binding model of a two-dimensional bipartite lattice, whose time-periodic driving protocol is illustrated in Fig.~2\textbf{a}. The driving protocol consists of four steps with equal duration of $T/4$. In each step, each site only couples to one of its four neighbouring sites (hence, the instantaneous system is dimerised).  A global dimerisation is introduced by letting the coupling strength in one of the four steps be different from other three. More specifically, the coupling strength is $\gamma$ for steps 1,2 and 4, while for step 3 the coupling strength is denoted by $\theta$. By varying $\gamma$ and $\theta$, a phase diagram that characterises all possible topological phases can be presented (see Supplementary Information). This Floquet system can exhibit two bandgaps, near quasienergies zero and $\pi$. We call these the `zero bandgap' and `$\pi$ bandgap', and the corner states in these bandgaps `zero modes' and `$\pi$ modes', respectively. Note that the concept of `zero modes' and `$\pi$ modes' is specific to Floquet systems and has no static counterpart.  The quasienergy bands have no polarisation or quadrupole moment, and are thus fundamentally different from all previous HOTIs (see Supplementary Information for the detailed topological characterisation).

We first consider the case $\theta = \gamma = 0.841 \pi$. The corresponding quasienergy spectrum is plotted in Fig.~2\textbf{b}. Here the $\pi$ bandgap is open (and can host corner states), whereas the zero bandgap is closed. To implement this 2D Floquet model, we use a 3D acoustic lattice with one axis ($z$) playing the role of time \cite{rechtsman2013}, as shown in Fig.~2\textbf{c}. In all the following calculations and demonstrations, we choose the lattice constant $a=20\sqrt{2}\,\textrm{mm}$, and four unit cells extend along both the $x$ and $y$ directions. Each site in the tight-binding model corresponds to a square air-hole waveguide with side length $l=10\,\textrm{mm}$ surrounded by hard acoustic boundaries. The coupling between two ajacent sites is accomplished by placing, between two adjacent square waveguides, a few thin connecting waveguides (these are not visible in Fig. 2\textbf{c}; see Supplementary Information for the design). By modulating the placement of the connecting waveguides along $z$, we realise an effective time-periodic driving following the protocol in Fig.~2\textbf{a}. In this construction, we take $L=336\,\textrm{mm}$ as the modulation period along $z$. The sample in Fig.~2\textbf{c} has length 3.5$L$. The coupling strength can be adjusted by altering the number of thin connecting waveguides. For example, by setting 12 connecting waveguides, $\theta = \gamma = 0.841 \pi$ can be satisfed at the acoustic frequency of 8000 Hz (see Supplementary Information for numerical calculation of coupling strength).

Now we demonstrate the dynamic properties of the $\pi$ modes. A speaker is placed at the lower left corner (indicated by a green arrow in Fig.~2\textbf{d}) to excite the corner states. The acoustic pressure at different propagation distances is recorded by a microphone (see Methods). Fig.~2\textbf{d} shows the measured evolution of the corner states, which exhibit strong localisation around the lower left corner. Notably, the intensity oscillates between the two sublattices near the corner, each taking half a period; this is a unique feature of the $\pi$ corner modes. These experimental observations are consistent with simulation results (see Supplementary Information and Methods for details about numerical simulations), thus verifying the existence of the $\pi$ corner modes.

We then further explore the properties of this lattice by taking different coupling strengths $\gamma$ and $\theta$.  We consider reducing the number of connecting waveguides in step 3 from 12 to 8, such that the coupling strength $\theta$ is reduced correspondingly to 0.568$\pi$ while $\gamma = 0.841 \pi$ is maintained at 8000 Hz. The resulting quasienergy spectrum is shown in Fig.~3\textbf{a}. In this case, both the $\pi$ bandgap and zero bandgap are open.  The numerically simulated eigenmode profiles in Figs.~3\textbf{b},\textbf{c} confirm the existence of zero modes and $\pi$ modes localised at the corners. The $\pi$ corner modes oscillate between two sublattices, consistent with Fig.~2\textbf{d}. Despite moderate changes in time, the zero modes mainly concentrate in one sublattice, similar to corner states in static HOTIs.

We then fabricated another experimental sample meeting the condition described in the previous paragraph ($\theta = 0.568\pi$ and $\gamma = 0.841 \pi$). The resulting dynamics, shown in Fig.~3\textbf{d}, is very different from the previously studied case that had only $\pi$ modes present.  Under corner excitation, the acoustic intensity is localised around the corner, but the mode profile does not repeat itself after one driving period, as is evident by comparing the acoustic intensities at $z/L=0$ and $z/L = 1$. Instead, a doubled period is observed by comparing acoustic intensities at $z/L=0$ and $z/L = 2$. The period doubling comes from the superposition of zero modes ($|0\rangle$) and $\pi$ modes ($|\pi\rangle$)---e.g., $a|0\rangle+b|\pi\rangle$, which evolves to another state after one driving period $U_{L}(a|0\rangle+b|\pi\rangle)=(a|0\rangle-b|\pi\rangle)$ and comes back to itself at two periods $U_{2L}(a|0\rangle+b|\pi\rangle)=(a|0\rangle+b|\pi\rangle)$, where $U_L$ is the evolution operator over one period \cite{bomantara2018}.  This unique period doubling feature has previously been predicted for discrete time crystals, and is a striking outcome of the coexistence of zero and and $\pi$ modes.

\begin{figure}
  \centering
  \includegraphics[width=\linewidth]{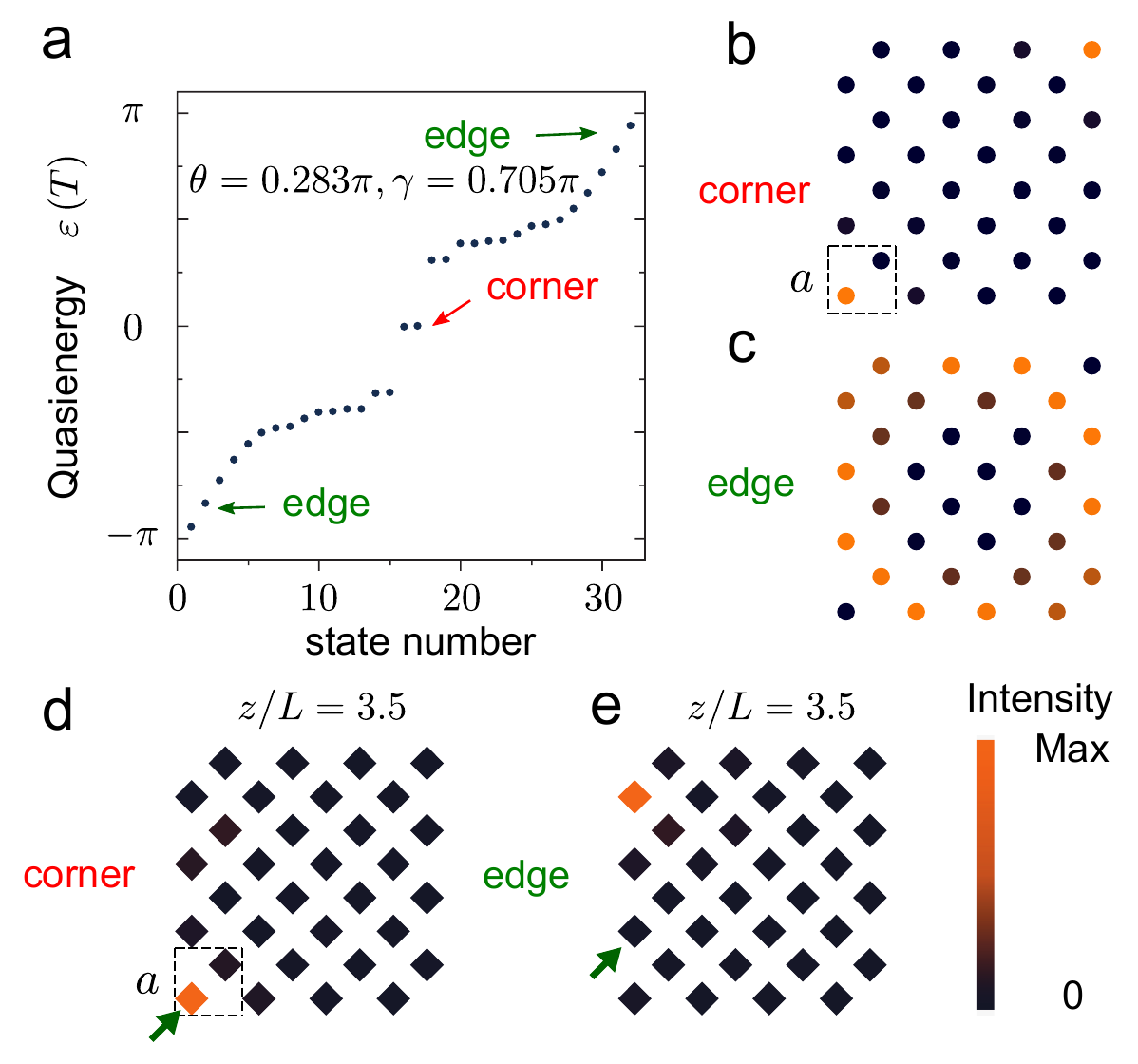}
  \caption{\textbf{Demonstration of coexisting corner states and chiral edge states.} \textbf{a}, Simulated quasienergy spectrum with $\gamma=0.705\pi$ and $\theta=0.283\pi$. \textbf{b}, Simulated everage eigenmode profile for the corner states. \textbf{c}, Simulated everage eigenmode profile for the chiral edge states. \textbf{d} Measured acoustic intensity distribution at $z/L = 3.5$ at 8000 Hz with corner excitation. The green arrow indicates the position of excitation at $z/L= 0$. \textbf{e}, Measured acoustic intensity distribution at $z/L = 3.5$ at 8000 Hz with edge excitation. The green arrow indicates the position of excitation at $z/L= 0$.}
  \label{fig4}
\end{figure}

The above zero and $\pi$ modes are protected by the higher-order topology associated with zero bandgap and $\pi$ bandgap, respectively. In fact, the topological properties of these two bandgaps can be controlled separately by tuning the coupling strengths $\gamma$ and $\theta$. Interestingly, it is possible to let one bandgap possess nontrivial first-order topology, while the other exhibits higher-order topology. In such a case, corner states and chiral edge states can simultaneously exist. (In all previously-studied static HOTIs, only the corner states were topologically protected, and any coexisting edge states were topologically trivial.) To study this phenomena, we reduce the number of connecting waveguides in driving steps 1, 2, and 4 to ten (so that $\gamma=0.705\pi$) and that in driving step 3 to 4 (so that $\theta=0.283\pi$). The simulated quasienergy spectrum in Fig.~4\textbf{a} shows that the zero bandgap now hosts zero corner modes (due to higher-order topology), while the $\pi$ bandgap is spanned by gapless chiral edge states (due to first-order topology). The calculated eigenmode profiles in Figs.~4\textbf{b},\textbf{c} confirm that these are indeed coexisting corner states and chiral edge states.

We fabricated a sample that meets these conditions ($\gamma=0.705\pi$ and $\theta=0.283\pi$). To probe the corner states and edge states separately, we conducted two measurements with different excitations, whose results are plotted in Figs.~4\textbf{d},\textbf{e}. In the first measurement (Fig.~4\textbf{d}), the source is placed at the lower left corner (indicated by the green arrow) at $z/L=0$. In this case, the corner state is excited and the acoustic intensity is found to be localised at the corner after an evolution of 3.5 driving periods. In the second measurement (Fig.~4\textbf{e}), the excitation occurs along the left edge (indicated by the green arrow) at $z/L = 0$.  The chiral edge state then propagates along the edge unidirectionally and moves up by around two lattice constants after an evolution of 3.5 driving periods. These observations provide direct evidences of the coexistence of corner states and chiral edge states.

In summary, we have proposed and experimentally demonstrated a Floquet HOTI in an acoustic lattice. Floquet higher-order topology exhibits unusual dynamic properties not found in static HOTIs. In particular, the coexistence of zero modes and $\pi$ modes may be interpreted as a realisation of an edge-state-based time crystal, whose robustness is tied to topological protection instead of many-body interactions \cite{bomantara2018}. These Floquet modes may also find applications in measurement-based quantum computing \cite{bomantara2020}, and the coexistence of corner states and chiral edge states may be useful for quantum state transfer \cite{Bomantara2020b}. Although the concept has been demonstrated on an acoustic platform, similar models can also be realised in photonic systems such as coupled ring resonators \cite{gao2016} and laser-written optical waveguides \cite{rechtsman2013,maczewsky2017,mukherjee2017} where the effects of non-Hermiticity and nonlinearity can be more easily studied, or even in a real time-dependent system \cite{darabi2020}.

\noindent{\large{\bf{Methods}}}

\textbf{Numerical simulation.} All full-wave simulations are performed by the finite element solver in Comsol Multiphysics (pressure acoustic module). The boundaries of the 3D printing materials (photosensitive resin) are modelled as rigid acoustic walls due to the large impedance mismatch with air (density $\rho=1.29$ $\text{kg/m}^3$ and sound speed $v=343$ m/s). In all simulations presented in the main text, the models have the same size as the ones used in experiment ($z/L=3.5$). The air boundaries at $z/L=0$ and $z/L=3.5$ are set to be radiation boundaries with a incident field applied to the lower left corner (Fig.~2\textbf{d}, Fig.~3\textbf{d} and Fig.~4\textbf{d}) or one site on the left edge (Fig.~4\textbf{f}) at $z/L=0$.

\textbf{Sample fabrication.} All samples are fabricated through a stereolithography apparatus with a resolution around 1 mm. In order to measure the acoustic intensity at different positions along $z$, each sample is devided into 6 pieces with cutting positions corresponding to $z/L=0,0.5,1,1.5,2,2.5,3$ and $3.5$. These small pieces are fabricated separately and then assembled togehter to construcut the sample.

\textbf{Experimental measurement.} All experiments are conducted using the same scheme. The acoustic wave is generated by a loudspeaker and guided into one lattice site at $z/L=0$ through a small tube. The output signals are record by a microphone (Br{\"u}el\&Kjaer Type 4182) that sweeps all the sites at the output plane. The measured signals are then processed by a analyser system (Br{\"u}el\&Kjaer 3160-A-022 module) to get the frequency-resolved spectrum. In all figures on experimental results, the data are normalised to the maximal value in the respective figure.

\end{document}